# Resonance structure in the γγ and $\pi^0\pi^0$ systems in dC interactions

Kh.U. Abraamyan[1,2]*, A.B. Anisimov[1], M.I. Baznat[3], K.K. Gudima[3,*], M.A. Kozhin[1], V.I. Kukulin[4,5], M.A. Nazarenko[6], S.G. Reznikov[1], and A.S. Sorin[7]

[1] VBLHEP JINR, 141980 Dubna, Moscow region, Russia
[2] International Center for Advanced Staudies, YSU, 0025, Yerevan, Armenia
[3] Institute of Applied Physics, MD-2028 Kishinev, Moldova
[4] Institute of Nuclear Physics, Moscow State University, 119992, Moscow, Russia
[5] Institut für Theoretische Physik, Universität Tübingen, Auf der Morgenstelle 14, D-72076 Tübingen, Germany
[6] Moscow State Institute of Radioengineering, Electronics and Automation, 119454 Moscow, Russia
[7] BLTP JINR, 141980 Dubna, Moscow region, Russia
* E-mail: abraam@sunhe.jinr.ru

**Abstract**

Along with $\pi^0$ and η mesons, a resonance structure in the invariant mass spectrum of two photons at $M_{\gamma\gamma} = 360 \pm 7 \pm 9$ MeV is observed in the reaction $d + C \to \gamma + \gamma + X$ at momentum 2.75 GeV/c per nucleon. Estimates of its width and production cross section are $\Gamma = 64 \pm 18$ MeV and $\sigma_{\gamma\gamma} = 98 \pm 24\,^{+93}_{-67}$ μb, respectively. The collected statistics amount to $2339 \pm 340$ events of $1.5 \cdot 10^6$ triggered interactions of a total number $\sim 10^{12}$ of $d$C-interactions. The results on observation of the resonance in the invariant mass spectra of two $\pi^0$ mesons are presented: the data obtained in the $d+C \to \gamma+\gamma$ reaction is confirmed by the $d+C \to \pi^0+\pi^0$ reaction: $M\pi\pi = 359.2 \pm 1.9$ MeV, $\Gamma = 48.9 \pm 4.9$ MeV; the ratio of $Br(R \to \gamma\gamma) / Br(R \to \pi^0\pi^0) = (1.8 \div 3.7) \cdot 10^{-3}$.

## 1. INTRODUCTION

Dynamics of the near-threshold production of lightest mesons and their interaction, especially ππ interaction, are of lasting interest. Two-photons decay of light mesons represents an important source of information. In particular, the γγ decay of light scalar mesons has been considered as a possible tool to deduce their nature. In this work the resonance structure with mass about 360 MeV in γγ and $\pi^0\pi^0$ systems is described.

## 2. EXPERIMENT

The data acquisition of production of neutral mesons and γ-quanta in $d$C interactions has been carried out with internal beams of the JINR Nuclotron [1 - 3]. The experimental setup is schematically represented in Fig. 1. The PHOTON-2 setup includes 32 γ-spectrometers of lead glass [4–6]. The modules of the γ-spectrometer are assembled into two arms of 16 units. These modules in each arm are divided into two groups of 8 units. The output signals in each group are summed up linearly and after discrimination by amplitude are used in fast triggering. In this experiment, the discriminator threshold was at the level of 0.4 GeV. Triggering takes place when there is a coincidence of signals from two or more groups from different arms.

## 3. EVENT SELECTION

The so-called event mixing method was used to estimate the combinatorial background: combinations of γ-quanta were sampled randomly from different events. For the general sampling and combinatorial background one, the same following selection criteria were used:
(1) the number of γ-quanta in an event, $N\gamma = 2$;
(2) the energies of γ-quanta, $E\gamma \geq 100$ MeV;
(3) the summed energy in real and random events $\leq 1.5$ GeV.
The invariant mass distributions before and after background subtraction are shown at Fig. 2 [2,3].

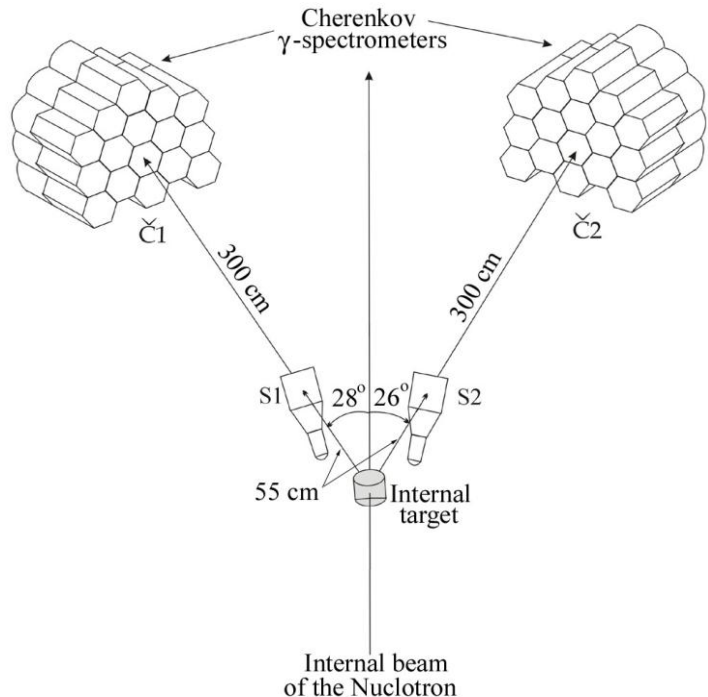

**Fig. 1.** The schematic drawing of the experimental PHOTON-2 setup. The $S1$ and $S2$ are scintillation counters.

## 4. DATA SIMULATION

To simulate $d+C$ reaction under question we use a transport code on the base Dubna cascade model [7] with upgrade elementary cross sections involved and with including experimental conditions.
The following γ-decay channels are taken into account: the direct decays of $\pi^0$, η, η' hadrons into two γ's, $\omega \to \pi^0\gamma$, $\Delta \to N\gamma$ and the Dalitz decay of $\eta \to \pi^+\pi^-\gamma$, $\eta \to \gamma e^+ + e^-$ and $\pi^0 \to \gamma e^+ + e^-$, the $\eta' \to \rho^0 + \gamma$, the $\Sigma \to \Lambda + \gamma$, the $\pi N$ and $NN$-bremsstrahlung. Furthermore the mechanism of η $3\pi^0$-decay [8] was checked. We simulated two channels of it: the direct decay into two photons $\eta \to 2\gamma$ and $\eta \to 3\pi^0$ which then decay into photons. Then the dibaryon mechanism of the two-photon emission [9] has been studied. The proposed mechanism $NN \to d_1^*\gamma \to NN\gamma\gamma$ proceeds trough a sequential emission of two photons, one of which is caused by production of the decoupled baryon resonance and the other is its subsequent decay.
These models reproduce quite accurately the observed η peak in the invariant mass distribution of γ pairs but there is no enhancement in the region where experimental data exhibit a resonance-like structure. Therefore we have included the additional channel of two-γ creation by two-pions interaction with the observable structure formation (which will be called below an R-resonance).

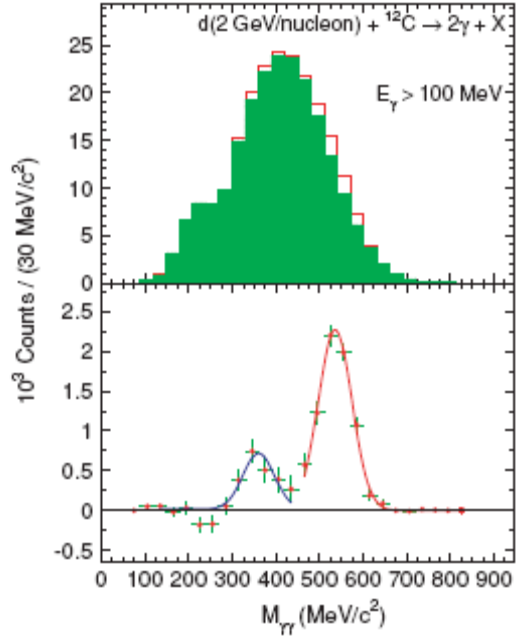

**Fig. 2.** Invariant mass distributions of γγ pairs satisfying criteria (1)–(3) without (upper panel) and with (bottom panel) the background subtraction for the reaction $d$ + C at 2.75 GeV/$c$ per nucleon. The curves are the Gaussian approximation of experimental points.

We assume that $R$-resonance can be created by ππ-interactions if the invariant mass of two pions obeys to Breight-Wigner distribution with observed parameters and both γ-quanta satisfy experimental condition. The two photons invariant mass spectrum and its comparison with the experiment is shown in Fig. 3.

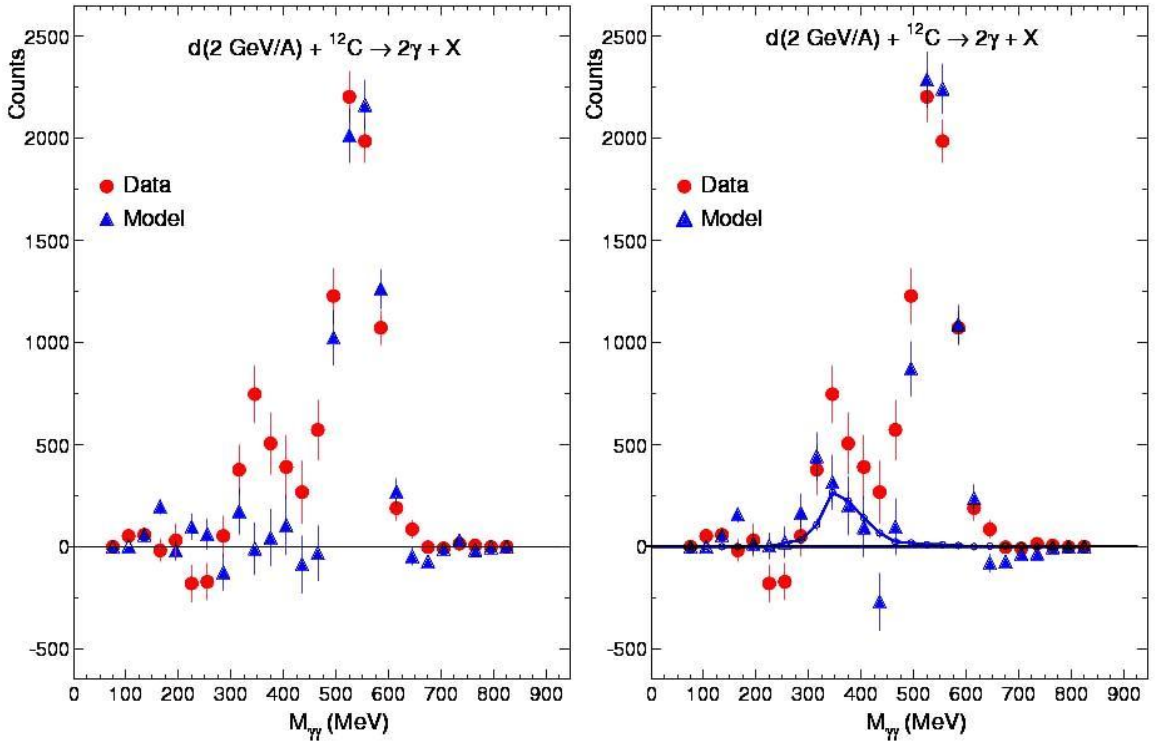

**Fig. 3.** The comparison of invariant mass distributions after background subtraction in $dC \rightarrow \gamma\gamma X$ to modeling result without (left panel) and with (right panel) including the $\pi\pi \rightarrow R \rightarrow \gamma\gamma$ channel.

Another more candidate for realization of dibaryon mechanism may be a model of the intermediate σ-dressed dibaryon [10]. This mechanism is now under investigation.

## 5. INVARIANT MASS SPECTRA OF TWO $\pi^0$ MESONS

We have analyzed the data for the reaction of $d + C$ with 4 and more detected photons. The detected photon multiplicities are shown in Fig. 4. For two $\pi^0$ identification we have formed and minimized the value:

$$S = (M_{ij} - m_\pi)^2 /(\Delta M_{ij})^2 + (M_{kl} - m_\pi)^2 /(\Delta M_{kl})^2, \qquad (1)$$

where $m_\pi$ = 135 MeV – the $\pi^0$ meson mass, $M_{ij}$, $M_{kl}$ – effective masses of different γγ combinations, $\Delta M_{ij}$, $\Delta M_{kl}$ – corresponding uncertainties in measured values of masses.

The invariant mass distributions of $\pi^0\pi^0$ pairs before and after background subtraction are shown in Fig. 5. The following selection criteria were used:

(i) the energies of photons, $E\gamma \geq 50$ MeV;
(ii) the energies of pions (the sum of two photon energy), $E\pi \geq 200$ MeV;
(iii) the sum of two pion energy in real and random events, $0.7 \leq E\pi\pi \leq 1.6$ GeV;
(iv) the opening angles of $\pi^0\pi^0$ pairs, $Cos\,\Theta\pi\pi \leq 0.93$.

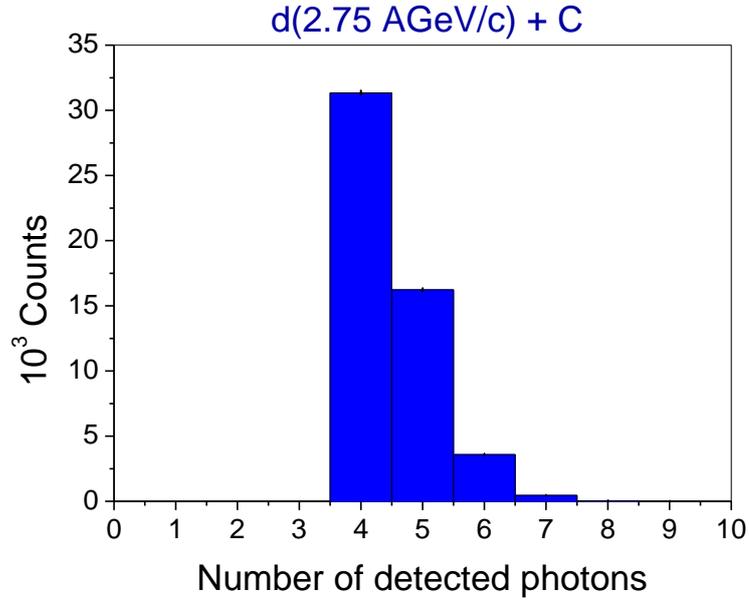

**Fig. 4.** The detected photon number distribution for the d+C reaction at 2.75 GeV/c per nucleon.

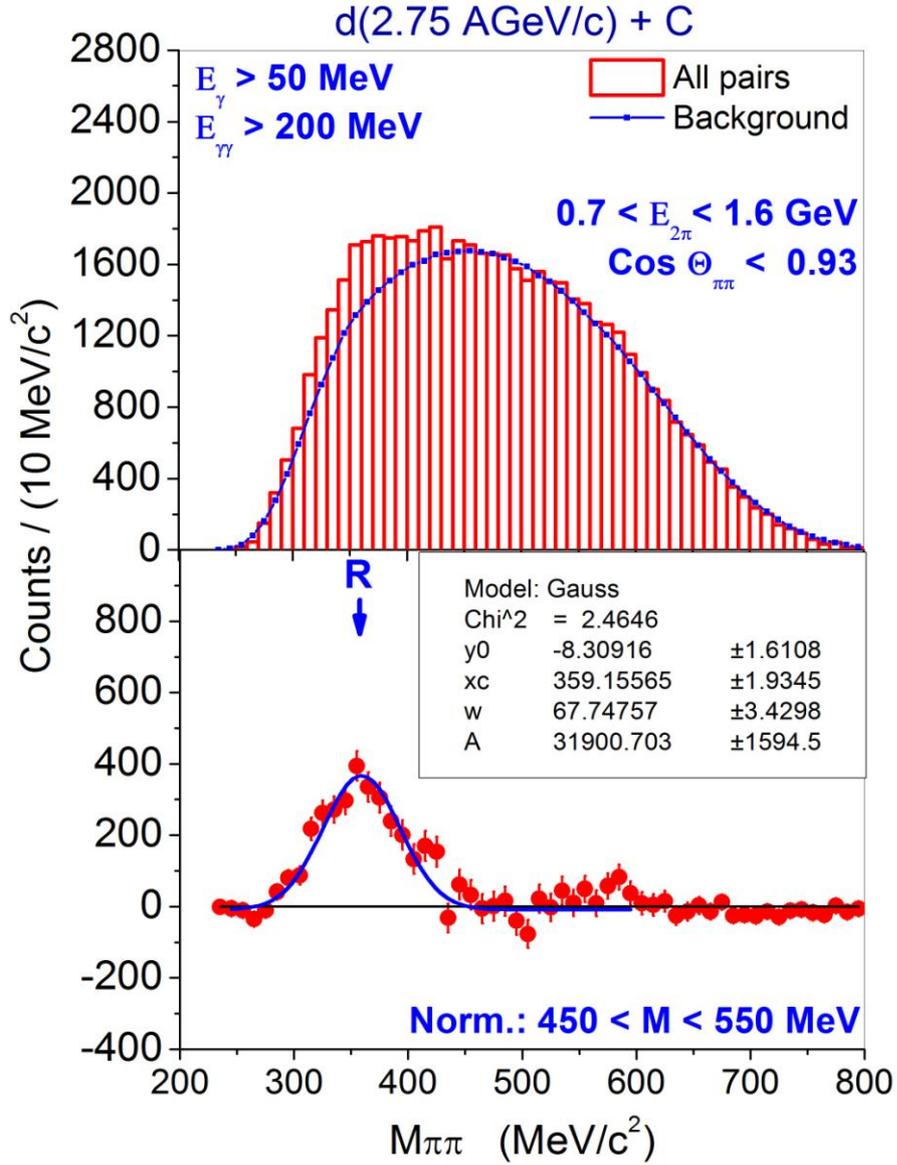

**Fig. 5.** The Invariant mass distributions of $\pi^0\pi^0$ pairs without (upper panel) and with (bottom panel) the background subtraction for the reaction $d$+C at 2.75 Gev/c per nucleon. The background are normalized to pair numbers in the interval: $450 < M\pi\pi < 550$ MeV/c$^2$.

## 6. CHECK THE OBSERVED PEAK

To elucidate the nature of the detected enhancement, we investigate the dependence of its position and width on:
   1) their energy selection level;
   2) the opening angle of two pions (see Fig. 6);
   3) the value of parameter $m_\pi$ in Eq. (1) (instead of the value of $m_\pi = 135$ MeV, see Fig. 7);
   4) the minimal value of S in Eq. (1) (see Fig. 8).

As is seen from Figs. 6-8, the result of the changing of observation conditions is quite robust: the position and width of the observed peak remain almost unchanged in different intervals of both energies and opening angles of $\pi^0$-mesons, namely, the mean peak position in the invariant mass distribution varies under different conditions in the range 355–368 MeV. The total number of detected events in the region 300-450 MeV (a summed number of pairs in the histogram in Fig. 5) after the background subtraction is $3099 \pm 152$.
.

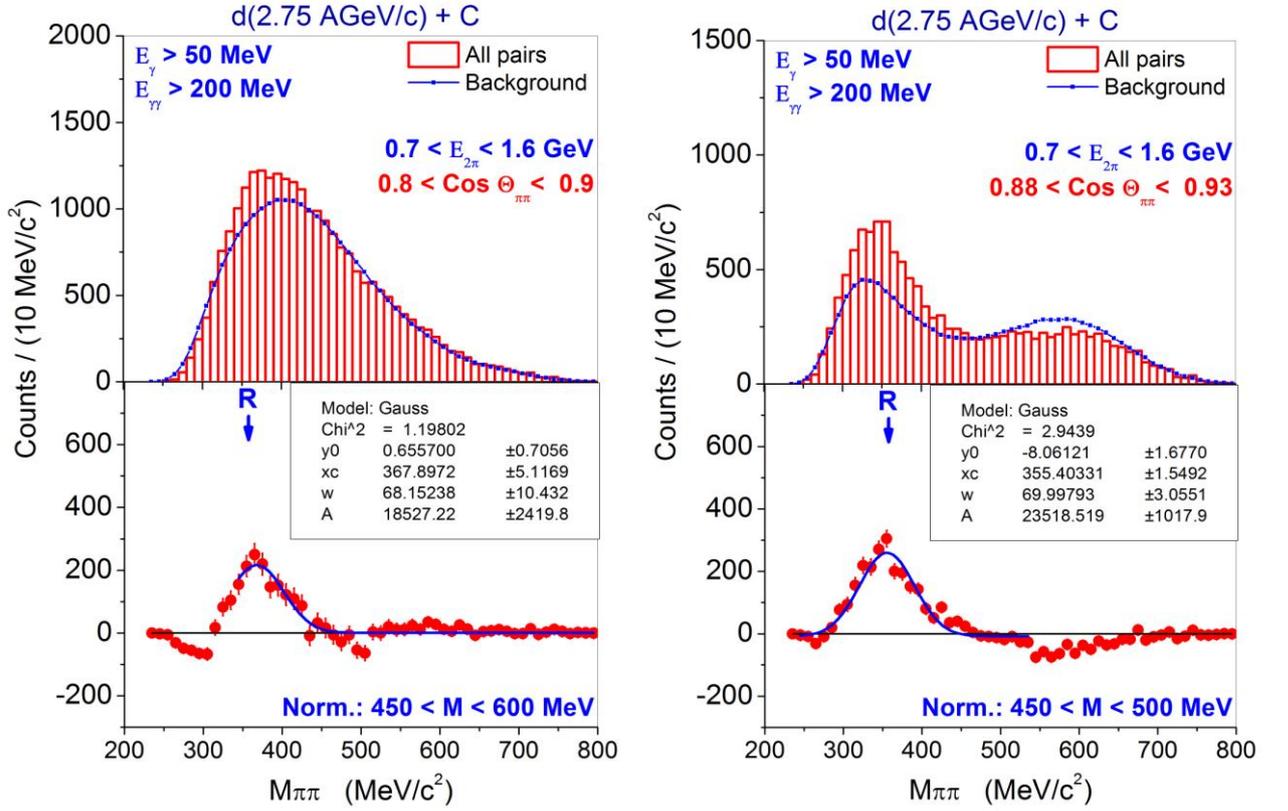

**Fig. 6.** The invariant mass distributions of two pions for the opening angles $0.8 < \cos\Theta_{\pi\pi} < 0.9$ (left) and $0.88 < \cos\Theta_{\pi\pi} < 0.93$ (right) under the selection criteria (i)-(iii).

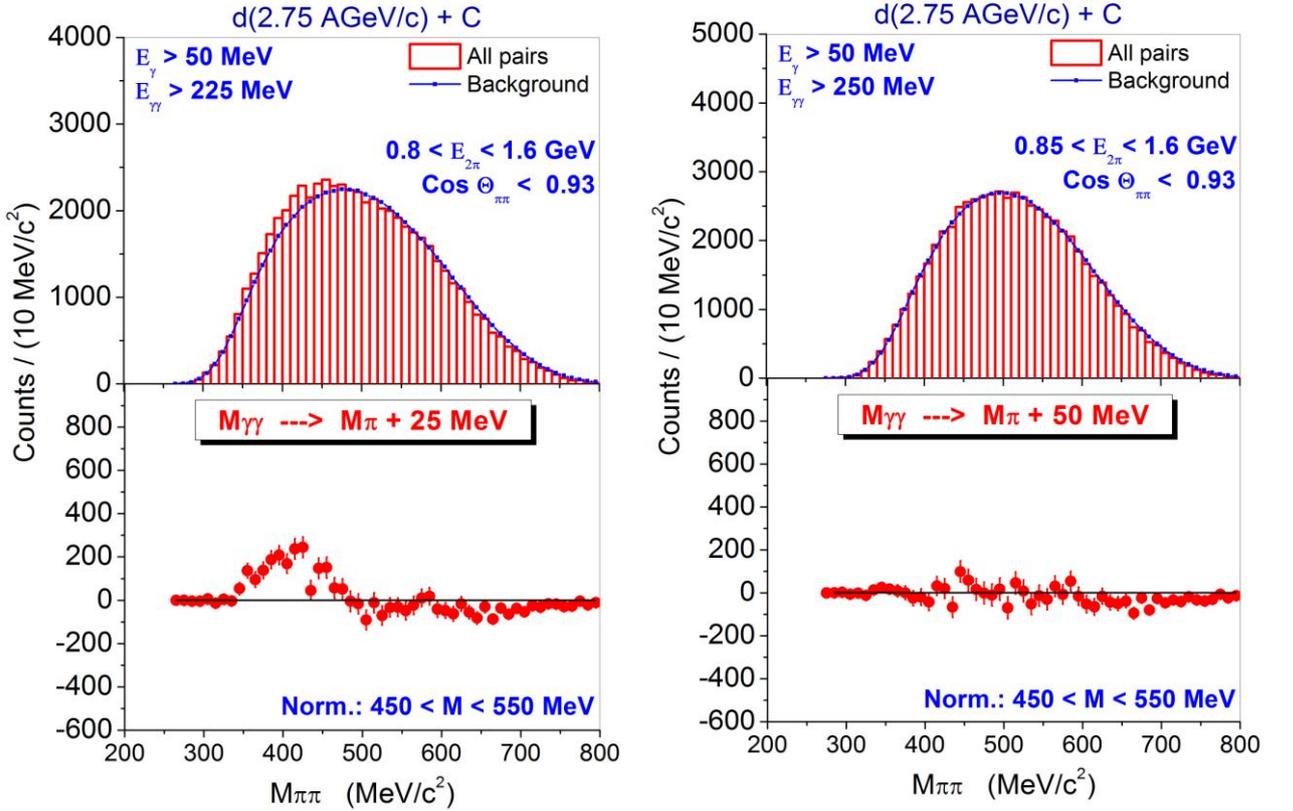

**Fig. 7.** The invariant mass distributions of two pairs of two photons at the mass $M_{\gamma\gamma} = m_\pi + 25$ MeV (left) and $M_{\gamma\gamma} = m_\pi + 50$ MeV (right) instead of $M_{\gamma\gamma} = m_\pi = 135$ MeV in Eq. (1).

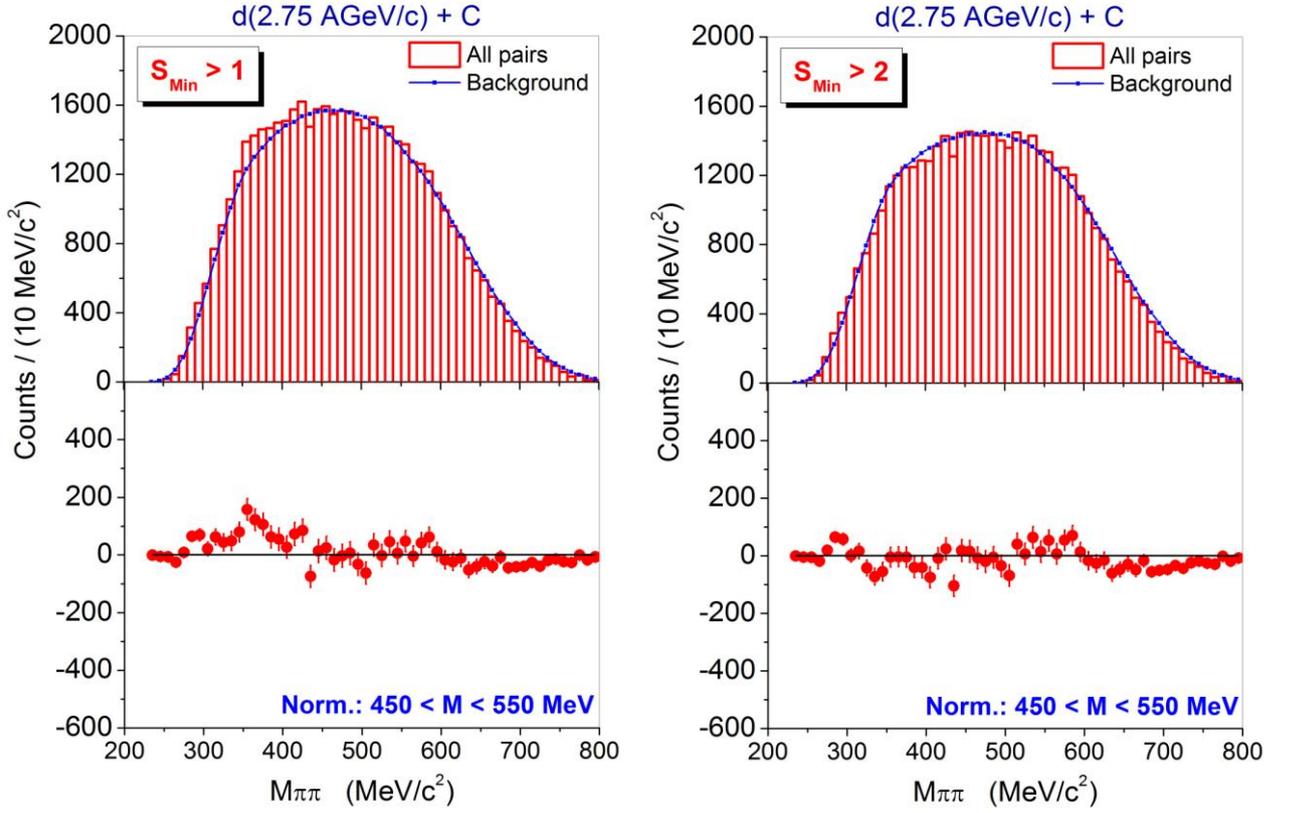

**Fig. 8.** The invariant mass distributions of two pions for two intervals of the minimal value of $S$ in Eq. (1).

## 7. THE MAIN CHARACTERISTICS OF THE OBSERVED RESONANCE

The observed peak in Fig. 5 is approximated by the Gaussian:

$$y = y_0 + \frac{A}{w\sqrt{\pi/2}} e^{-2\frac{(x-xc)^2}{w^2}}. \qquad (2)$$

The mean values of the peak measured parameters are:
$$M\pi\pi = xc = 359.2 \pm 1.9 \text{ MeV and } 2\sigma_{measur} = w = 67.8 \pm 3.4 \text{ MeV}.$$

The internal width of the observed resonance is defined by the measured width $w$ and also specified by the spectrometer resolution $w_{sp}$:

$$w_{int} = (w^2 - w_{sp}^2)^{1/2}. \qquad (3)$$

The spectrometer resolution for the resonance invariant mass range is:
$$w_{sp} (340 < M < 360 \text{ MeV}) \approx 47 \text{ MeV}.$$

The value of $w$ in the Gaussian distribution (2) practically coincides with the width $\varGamma$ in the Breight-Wigner function (see Fig. 9); thus, according to Eq. (3), the intrinsic width of the observed resonance structure is:
$$\varGamma \approx w_{int} = 48.9 \pm 4.9 \text{ MeV}.$$

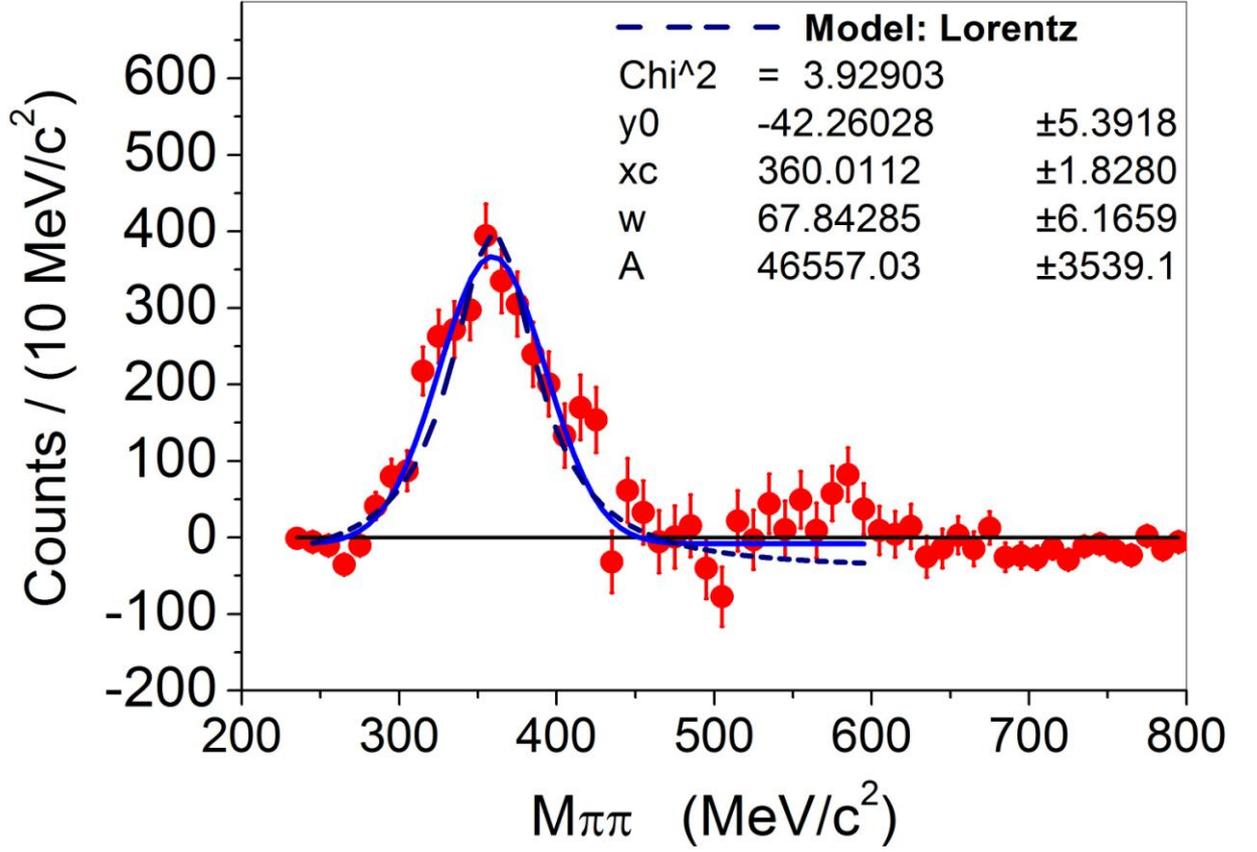

**Fig. 9.** Approximation of the peak in Fig. 5 by the Breight-Wigner function (dashed line). The solid line is the approximation by the Gaussian.

## 8. ESTIMATE OF BRANCHING RATIO OF THE $R \to \pi^0\pi^0$ DECAY

To define the $R \to \pi^0\pi^0$ events detection efficiency we have simulated about $1.4 \cdot 10^{10}$ $d$C-interactions. The $R$-resonance was included by above described method (see Item 4). If the $R$ resonance has been formed, it is assumed to decay only in two $\pi^0$ mesons. Each simulated event was turned in the $\Phi$ plane by $\Delta\Phi = 22.5°$ (7 times).

The invariant mass distributions of $\pi^0\pi^0$ pairs selected from above described procedure (see Item 5) are shown in Fig. 10.

The branching ratio of the $R \to \pi^0\pi^0$ decay we have estimated by formula:

$$\frac{Br(R \to \pi^0\pi^0)}{Br(R \to \gamma\gamma)} = \frac{N(R \to \pi^0\pi^0)/\varepsilon(R \to \pi^0\pi^0)}{N(R \to \gamma\gamma)/\varepsilon(R \to \gamma\gamma)}, \quad (4)$$

where $N$ – is the number of selected events in the experiment,

$$\varepsilon = N^{Mod}_{Select}(R) / N^{Mod}_{All}(R)$$

– is the detection and selection efficiency, $N^{Mod}_{Select}(R)$ and $N^{Mod}_{Select}(R)$ are the selected and the all numbers of $R$-resonances in the model, respectively.

The $R \to \pi^0\pi^0$ events detection efficiency in the real conditions of the experiment we have estimated by extrapolation of data obtained at different values of threshold propagation factor (see Fig. 11).

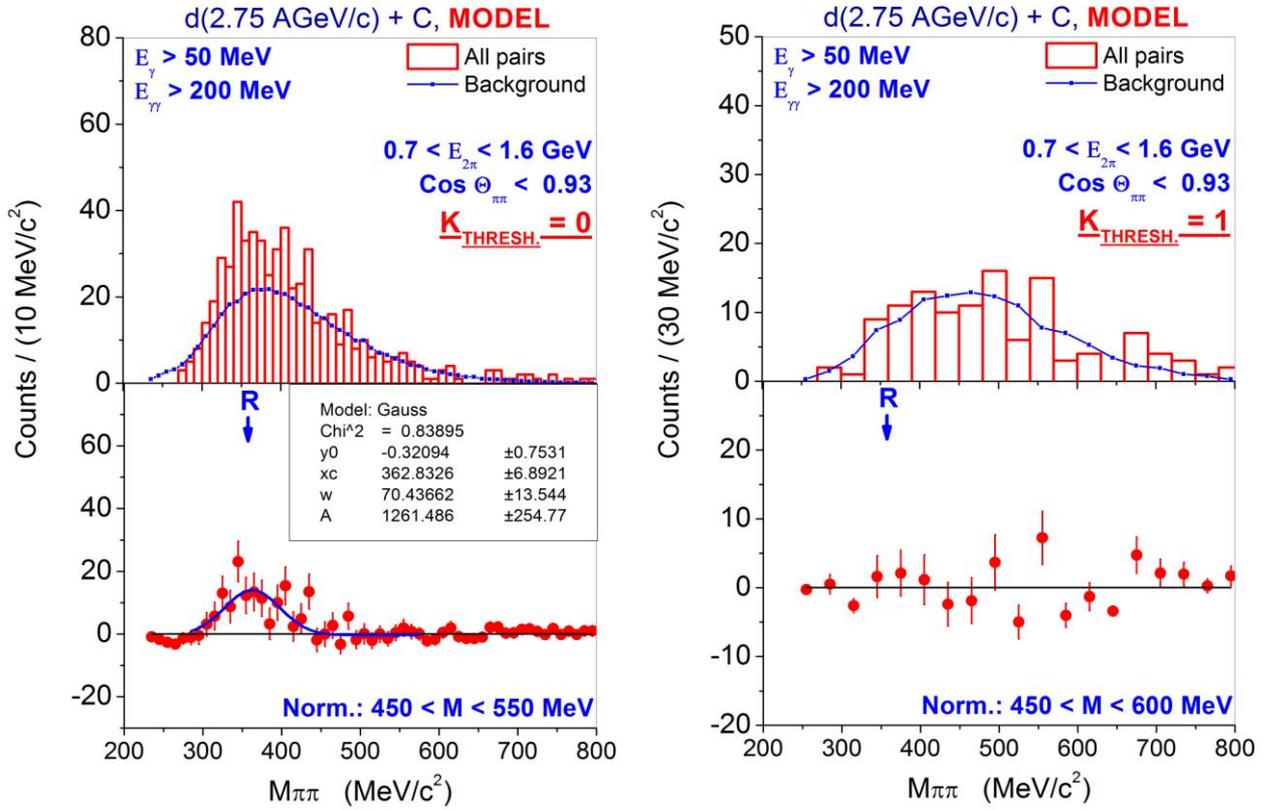

**Fig. 10.** The Invariant mass distributions of $\pi^0\pi^0$ pairs in the same conditions as in Fig. 5, at the discriminator threshold equals to 0 (left) and in the real conditions of the experiment (right).

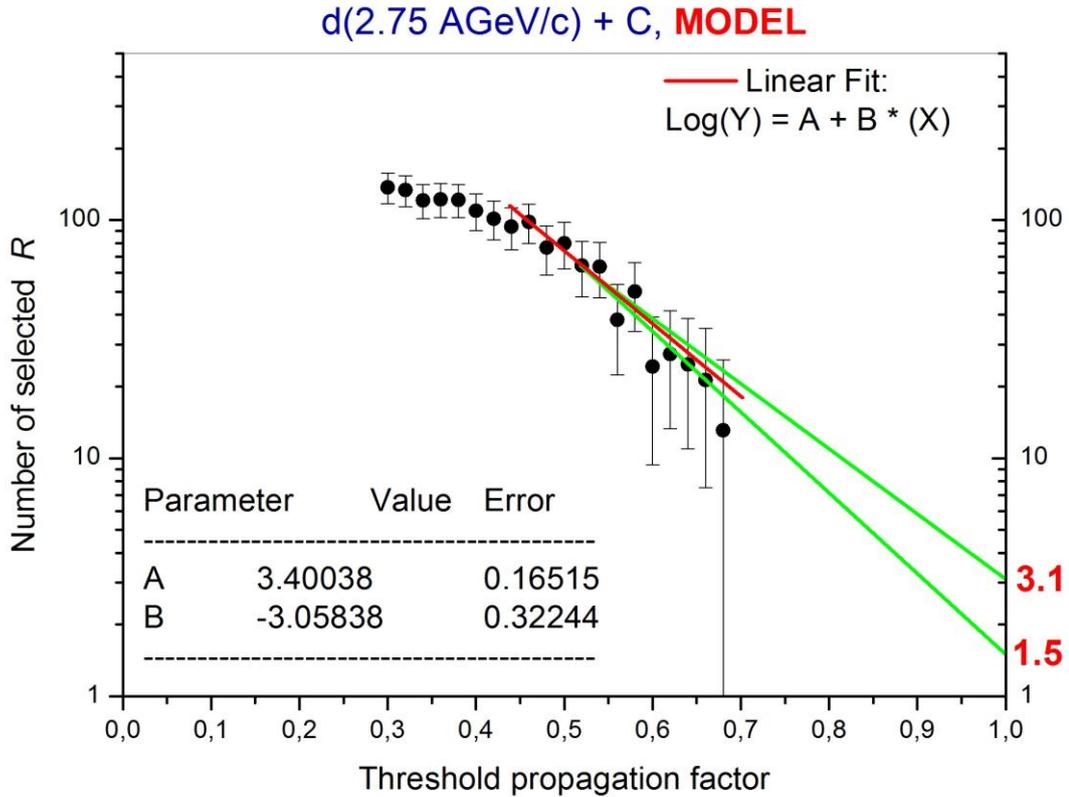

**Fig. 11.** The number of selected $R \to \pi^0\pi^0$ events (the total number of simulated events in the region 300–450 MeV after the background subtraction) versus of the discriminator threshold propagation factor.

From Eq. (4) we have estimated the branching ratio of $R \to \pi^0\pi^0$ decay:
$$Br(R \to \pi^0\pi^0) = (0.27 \div 0.57) \cdot 10^3 \, Br(R \to \gamma\gamma).$$

## 9. CONCLUSION

1. Following an experiment at the JINR Nuclotron the resonance-like enhancement was observed in two-gamma spectrum in $d$C-interactions (2.75 GeV/c per nucleon). Estimates of its characteristics are: $M\gamma\gamma$ = 360 ± 7 ± 9 MeV, $\Gamma$ = 64 ± 18 MeV, $\sigma_{\gamma\gamma}$ = 98 ± 24 $^{+93}_{-67}$ μb.
2. The data obtained in the $d+C \to \gamma+\gamma$ reaction is confirmed by the $d+C \to \pi^0+\pi^0$ reaction: $M\pi\pi$ = 359.2 ± 1.9 MeV, $\Gamma$ = 48.9 ± 4.9 MeV.
3. The ratio of $Br(R\to\gamma\gamma) / Br(R\to\pi^0\pi^0) = (1.8 \div 3.7)\cdot 10^{-3}$.

## ACKNOWLEDGMENTS

This work is supported in part by the RFBR Grant No. 11-02-01538-a.

REFERENCES
1. Kh. U. Abraamyan, A. N. Sissakian, and A. S. Sorin, "Observation of New Resonance Structure in the Invariant Mass Spectrum of Two Gamma-Quanta in dC-Interactions at Momentum 2.75 GeV/*c* per Nucleon", arXiv:nucl-ex/0607027.
2. Kh. U. Abraamyan et al., Phys. Rev. C **80**, 034001 (2009).
3. Kh. U. Abraamyan et al., Doklady Physics, **55**, 161 (2010).
4. Kh. U. Abraamyan et al., Phys. Lett. B **323**, 1 (1994); Yad. Fiz. **59**, 271 (1996) [Phys. At. Nucl. **59**, 252 (1996)]; Yad. Fiz. **60**, 2014 (1997) [Phys. At. Nucl. **60**, 1843 (1997)]; Yad. Fiz. **68**, 1020 (2005) [Phys. At. Nucl. **68**, 982 (2005)].
5. Kh. U. Abraamyan et al., Prib. Tekh. Eksp. **1**, 57 (1989); Prib. Tekh. Eksp., No. 6, 5 (1996) [Instrum. Exp. Tech. **39**, 775 (1996)].
6. M. N. Khachaturian, JINR Commun. E1_85_55 (Dubna, 1985); R. G. Astvatsaturov et al., Nucl. Instr. Methods Phys. Res. **163**, 343 (1979).
7. K. K. Gudima and V. D. Toneev, Nucl. Phys. A **400**, 173c (1983); K. K. Gudima et al., Nucl. Phys. A **401**, 329 (1983).
8. L. M. Brown and P. Singer, Phys. Rev. Lett. **8**, 460 (1962).
9. A. S. Khrikin et al., Phys. Rev. C **64**, 034002 (2001); Nucl. Phys. A **721**, 625c (2003).
10. V. I. Kukulin, V. N. Pomerantsev, M. Kaskulov, and A. Faessler, J. Phys. G **30**, 287 (2004); J. Phys. G**30**, 309 (2004).